\documentclass[a4paper,12pt]{article}
\usepackage{epsfig}
\usepackage{epstopdf}
\usepackage[pdftex]{color}
\usepackage[sumlimits,intlimits,namelimits]{amsmath} 

\usepackage[english]{babel}

\usepackage{graphicx}

\usepackage{geometry}
\numberwithin{equation}{section}
\begin{document}
\begin{titlepage}
\begin{flushright}
SI-HEP-2017-01 \\
QFET-2017-01 \\[0.2cm]
%\today
\end{flushright}

\vspace{1.2cm}
\begin{center}
{\Large\bf 
Inclusive Semitauonic \boldmath $B$  Decays to Order 
${\cal O} (\Lambda_{QCD}^3/m_b^3)$ \unboldmath}
\end{center}

\vspace{0.5cm}
\begin{center}
{\large Thomas Mannel$^1$, Aleksey V. Rusov$^{1,2}$ and Farnoush Shahriaran$^1$} \\[0.5cm]
\textsl{$^1$Theoretische Physik 1, Naturwissenschaftlich-Technische Fakult\"at,  \\ 
Universit\"at Siegen, D-57068 Siegen, Germany}
\end{center}

\begin{center}
\textsl{%
$^2$ Department of Theoretical Physics,
P.G. Demidov Yaroslavl State University, \\
150000, Yaroslavl, Russia}
\vspace*{0.8cm}
\end{center}

\vspace{0.8cm}
\begin{abstract}
\vspace{0.2cm}\noindent
We calculate the decay width and the $\tau$-lepton energy distribution
as well as relevant moments for inclusive $\bar B \to X_c \tau \bar \nu_\tau$ process 
including power corrections up to order $\Lambda_{QCD}^3/m_b^3$
and QCD corrections to the partonic level.
We compare the result with the sum of the standard-model  predictions of the branching fractions 
of the exclusive semileptonic $\bar B \to (D, D^*, D^{**}) \tau \bar \nu_\tau$ decays
as well as with the relevant experimental data.
Our prediction is in agreement with the LEP measurement and is consistent with 
the standard-model calculation of the exclusive modes. We discuss the impact from 
physics beyond the Standard Model. 
\end{abstract}
\hspace*{10mm}{\bf Keywords}: Heavy quark physics; Heavy quark expansion; Inclusive decays

\end{titlepage}

%papercontent
\newpage
\pagenumbering{arabic}
\section{Introduction}
Semi-tauonic $B$ decays have attracted renewed attention after the measurements 
of the exclusive channels $\bar B \to D^{(*)} \tau \bar{\nu}$, which exhibit a tension with the
%predictions of the 
Standard Model (SM) \cite{BaBarOld, BaBar2, LHCb, BelleNew}.
In fact, the theoretical predictions within the SM
turn out to be quite precise, since one of the relevant form factors can be inferred from 
the decays into light leptons (electrons and muons), while the longitudinal form factor that appears 
only for the heavy $\tau$ lepton can be related to the known one by heavy quark 
symmetries (HQS). Although the use of HQS implies corrections of the order 
$\Lambda_{\rm QCD} / m_c$, a good precision is maintained due to the fact, that the  
contribution of the longitudinal form factor receives an additional suppression factor 
$m_\tau^2 / m_B^2$. 

However, there is another problem with the current data on exclusive semi-tauonic 
$B$ decays which is related to the degree of saturation of the inclusive 
$\bar B \to X_c \tau \bar{\nu}$  rate. There is on the one hand a measurement of this inclusive 
rate based on the $b$-hadron admixture as it was generated by LEP \cite{PDG}, 
$$
{\rm Br}(b\mbox{-admix} \to X \tau \bar{\nu})  = (2.41 \pm 0.23)\%
$$
which to leading order in the heavy quark expansion (HQE) should be the branching 
ratio for each individual hadron.  
On the other hand, one may also compute the inclusive semi-tauonic ratio
$R(X_c) = \Gamma(\bar B \to X_c \tau \bar{\nu}) / \Gamma(\bar B \to X_c \ell \bar{\nu})$,
where $\ell$ is a light lepton. This ratio does not depend on $V_{cb}$ and can
been computed within the HQE very precisely. In combination with the accurately
measured branching ratio ${\rm Br} (\bar B \to X_c \ell \bar{\nu})$ one finds
in the $1S$ scheme including corrections up to $1/m_b^2$ \cite{LigetiN2}
$$
{\rm Br}(B^- \to X_c \tau \bar{\nu}) = (2.42 \pm 0.05)\% 
$$
in full agreement with the LEP measurement. Taking the current data for $R(D)$ and 
$R(D^*)$ at face value, the two exclusive decay modes $\bar B \to D \tau \bar{\nu}$ and 
$\bar B \to D^* \tau \bar{\nu}$ would at least fully saturate 
(if not oversaturate) the inclusive rate.  

This situation has motivated us to perform an independent calculation of 
$\bar B \to X_c \tau \bar{\nu}$ within the HQE. 
The calculation presented in \cite{LigetiN1} makes use of the $1S$ scheme and 
includes terms up to order $1/m_b^2$. In this paper we present a 
calculation in the kinetic scheme and include terms up to order $1/m_b^3$, 
improving the existing calculations by including the next order in the HQE. 
In the light of the quite precise prediction for the inclusive 
$\bar B \to X_c \tau \bar{\nu}$ rate we discuss the 
theoretical predictions for the exclusive channels 
$\bar B \to D^{(*,**)} \tau \bar{\nu}$ and compare to the 
current experimental situation.

\section{The Inclusive $\bar B \to X_c \tau \bar{\nu}$ Decay}  
\subsection{Outline of the Calculation}
The matrix element for the $\bar B \to X_c \ell \bar \nu$ $(\ell = e, \mu, \tau)$ 
decay can be written in terms of the low energy effective Hamiltonian for the weak process $b\to c \ell \bar \nu$:
\begin{equation}
{\cal H}_W = \frac{G_F V_{cb}}{\sqrt{2}}J_{L}^\alpha J_{H\alpha}^{\phantom{\alpha}} 
+{\rm h. c.},
\label{eq:weak-Hamiltonian}
\end{equation}
where 
$J^\alpha_L = \bar \ell \gamma^\alpha(1-\gamma^5)\nu$ 
and $J^\alpha_H =\bar{c}\gamma^\alpha (1-\gamma^5) b$ are the leptonic and hadronic currents, respectively, and $V_{cb}$ is the CKM matrix element involved in the decay. 

We express the triple-differential distribution for $\bar B \to X_c \ell \bar \nu$ 
in terms of the energies of the lepton and neutrino $E_\ell$ and $E_\nu$ and the 
dilepton invariant mass $q^2 = (p_\ell + p_{\nu})^2$ as 
\begin{equation}
\frac{d\Gamma}{dE_\ell dq^2 dE_\nu} = \frac{G_F^2|V_{cb}|^2}{16 \pi^3}
L_{\alpha\beta} W^{\alpha\beta},
\label{eq:triple-distr}
\end{equation}
where $L_{\alpha\beta}$ and $W_{\alpha\beta}$ 
are called the leptonic and hadronic tensors, respectively.
In the Standard Model, the leptonic tensor takes the form 
\begin{equation}
L^{\alpha\beta}=\sum_{\rm lepton\,spin}\langle 0 |J^{\dagger\alpha}_{L}|\ell \bar \nu \rangle 
\langle \ell \bar \nu | J^\beta_{L}|0\rangle\\
=8 (p_\ell^\alpha p_{\nu}^\beta + p_\ell^\beta p_{\nu}^\alpha - 
g^{\alpha\beta} (p_\ell \cdot p_{\nu} ) -
i \varepsilon^{\rho\alpha\sigma\beta}p_{\ell_\rho} p_{\nu_{\sigma}})
\label{eq:LT}
\end{equation}
and the hadronic tensor is defined as
\begin{equation}
W^{\alpha\beta}= \frac 1 4 \sum_{X_c} \frac{1}{2 m_B}(2\pi)^3\langle \bar B |J^{\dagger \alpha}_{H}|X_c\rangle \langle X_c|J^\beta_{H}| \bar B \rangle \delta^{(4)} (p_B - q - p_{X_c}).
\label{eq:HT}
\end{equation}
Its general decomposition into scalar functions 
$W_j = W_j ((v\cdot q), q^2) $, $j = 1, \cdots, 5$ reads
\begin{equation}
W^{\alpha\beta} = - g^{\alpha\beta} W_1 + v^\alpha v^\beta W_2 - 
i \epsilon^{\alpha\beta\rho\sigma} v_\rho q_\sigma W_3 + q^\alpha q^\beta W_4
+(q^\alpha v^\beta + q^\beta v^\alpha) W_5 \, .
\end{equation}
After contraction of leptonic and hadronic tensors the triple
differential decay rate takes the form:  
\begin{eqnarray}
\frac{d\Gamma}{dE_\ell \, dq^2 dE_\nu} & = & \frac{G_F^2|V_{cb}|^2}{2 \pi^3} 
\Big\{q^2 \, W_1 + \Big(2E_\ell E_\nu-\frac{q^2}{2}\Big) W_2 
+ q^2 (E_\ell-E_\nu) W_3  
\label{eq:trip-distr-contr} \\
& + & \frac{1}{2} \,  m_{\ell }^2 \left[-2 \, W_1 + W_2 - 2 \left(E_{\nu }+E_{\ell }\right) W_3 
+ q^2 \, W_4 + 4 E_{\nu } W_5 \right] - \frac{1}{2} m_\ell^4 W_4 \Big\}.
\nonumber
\end{eqnarray} 
Due to the optical theorem, the hadronic tensor $W^{\alpha \beta}$ is related to the discontinuity 
of a time-ordered product of currents:
\begin{equation}
T^{\alpha \beta} = - \frac{i}{4}  \int d^4 x \, e^{-i q x}
\frac{\langle \bar B | {\rm T} \left\{ J_H^{\dagger \alpha} (x) 
J_H^\beta (0) \right\} | \bar B \rangle}
{2 m_B}
\end{equation}
via the relations
\begin{equation}
- \frac{1}{\pi} {\rm Im} T_j = W_j
\end{equation}
with the structure functions $T_i$ defined  in analogy to $W^{\alpha \beta}$:
\begin{equation}
T^{\alpha\beta} = - g^{\alpha\beta} T_1 + v^\alpha v^\beta T_2 - 
i \epsilon^{\alpha\beta\rho\sigma} v_\rho q_\sigma T_3 + q^\alpha q^\beta T_4
+(q^\alpha v^\beta + q^\beta v^\alpha) T_5 \, . 
\end{equation}

Inserting $p_b = m_b v + k$ for the momentum of the $b$ quark and expanding in the 
residual momentum $k \sim {\cal O} (\Lambda_{\rm QCD})$ yields the standard OPE as 
it is used for the light leptons. 
A simple way to derive this OPE at tree level based on an external-field method 
has been derived in \cite{sasha}.  

In order to calculate $\tau$-lepton energy spectrum and decay width 
we need to define the kinematic boundaries
of the variable involved in the triple differential decay rate (\ref{eq:trip-distr-contr}).
We introduce the following dimensionless variables 
\begin{equation}
\hat{q}^2=\frac{q^2}{m_b^2},\quad\quad x=\frac{2E_\nu}{m_b},\quad\quad y=\frac{2E_\tau}{m_b}
\end{equation}
and the mass parameters
\begin{equation}
\rho=\frac{m_c^2}{m_b^2},\quad\quad \eta=\frac{m_\tau^2}{m_b^2}.
\end{equation}
We first perform an integration over the energy of the final state neutrino $E_\nu$
and in terms of corresponding dimensionless variable $x$
the limits of integration are determined as
\begin{equation}
\frac{\hat{q}^2-\eta}{y_+} \le x  \le \frac{\hat{q}^2-\eta}{y_-}, 
\quad  \quad y_\pm=\frac{1}{2}\left(y\pm\sqrt{y^2-4\eta}\right).
\label{eq:x-lim-int}
\end{equation}
Subsequently we perform the integration over variable $\hat q^2$ with 
corresponding boundaries:
\begin{equation}
y_-\left(1-\frac{\rho}{1-y_-}\right) \le \hat{q}^2 \le y_+\left(1-\frac{\rho}{1-y_+}\right),
\label{qsq-lim-int}
\end{equation}
and one gets the $\tau$-lepton energy distribution.
Integration over all possible values of the $\tau$-lepton energy
\begin{equation}
2\sqrt{\eta} \le y \le 1+\eta-\rho
\label{y-lim-int}
\end{equation}
allows us to calculate decay width.

In this way we obtain the analytic result for the decay width which can 
be presented in the following form:
\begin{equation}
\Gamma( \bar B \to X_c \tau \bar \nu) =  \Gamma_0 \, (1+A_{ew}) \!\!
\left[C_0^{(0)} + \frac{\alpha_s}{\pi} C_0^{(1)}
+ C_{\mu_\pi^2} \cfrac{\mu_\pi^2}{m_b^2} + C _{\mu_G^2} \cfrac{\mu_G^2}{m_b^2}
+ C_{\rho_D^3}\cfrac{\rho_D^3}{m_b^3} + C_{\rho_{LS}^3}\cfrac{\rho_{LS}^3}{m_b^3} \right], 
\label{eq:decay-width-res}
\end{equation}
where nonperturbative parameters $\mu_\pi^2, \mu_G^2, \rho_D^3, \rho_{LS}^3$ are defined as:
\begin{eqnarray}
2 m_B \, \mu_\pi^2 & = & 
- \langle B(p) |\bar b_v (i D)^2 b_v| B(p) \rangle,
\\
2 m_B \, \mu_G^2 & = & 
\langle B(p) |\bar b_v (i D_\mu) (i D_\nu) (- i \sigma^{\mu \nu}) b_v| B(p) \rangle,
\\
2 m_B \, \rho_D^3 & = & 
\langle B(p) |\bar b_v (i D_\mu)(i v \cdot D) (i D^\mu) b_v| B(p) \rangle,
\\
2 m_B \, \rho_{LS}^3 & = & 
\langle B(p) |\bar b_v (i D_\mu)(i v \cdot D) (i D_\nu)(-i\sigma^{\mu \nu}) b_v| B(p) \rangle \, .
\end{eqnarray}
Note that this corresponds to a ``covariant'' definition of these parameters 
using the full covariant derivatives instead of only their spatial components,
for a more detailed discussion see~\cite{sasha}.  

The coefficients $C_0^{(0)}, C_{0}^{(1)}, C_{\mu_\pi^2}, C_{\mu_G^2}, C_{\rho_D^3}, C_{\rho_{LS}^3}$
depend on $\rho$ and $\eta$,  and we define
\begin{equation}
\Gamma_0 = \frac{G_F^2 |V_{cb}|^2 \, m_{b}^5}{192\pi^3}.
\end{equation}
The calculation of the decay width revealed that - as in the case 
of a massless lepton  - the corresponding
coefficients $C_{\rho_{LS}^3}$ for massive $\tau$-lepton also vanishes, $C_{\rho_{LS}^3} = 0$. 
The explicit analytic expressions for coefficients 
$C_0^{(0)}, C_{\mu_\pi^2}, C_{\mu_G^2}, C_{\rho_D^3}$
as functions of $\rho$ and $\eta$ can be found in Appendix.
The derived expressions for $C_0^{(0)}, C_{\mu_\pi^2}$ and $C_{\mu_G^2}$ 
are in agreement with the corresponding results of \cite{nir, Balk},
while analytic formula for $C_{\rho_D^3}$ represents a new result of this paper which
in the particular case $m_\ell \to 0$ (or equivalently $\eta \to 0$) 
reproduces the corresponding expression in~\cite{sasha}
originally derived in \cite{Gremm:1996df}.
Moreover, we include perturbative radiative corrections to the partonic level
of the decay width using results of \cite{Jezabek:1996}.  
This correction is presented as $C_0^{(1)}$ in eq.~(\ref{eq:decay-width-res}).
Additionally, we include the electroweak correction $A_{\rm ew}$ to the decay width 
which is well-known and can be found in \cite{sirlin}:
\begin{equation}
1+A_{\rm ew}\approx\Big(1+ \frac{\alpha_{\rm em}}{\pi} \, 
\ln \frac{M_Z}{m_b}\Big)^2\approx\,1.014.
\end{equation}

Moreover, we calculate the $\tau$-lepton energy distribution and their moments.
We define the moments of the $\tau$-lepton energy distribution as in \cite{schwanda}
\begin{equation}
\label{eq:lept-energy-mom-def}
M_\tau^n \equiv \langle E_\tau^n \rangle_{E_\tau > E_{\rm cut}} = 
\frac{\int_{E_{\rm cut}}^{E_{\rm max}} 
d E_\tau \, E_\tau^n \displaystyle\frac{d \Gamma}{d E_\tau}}
{\int_{E_{\rm cut}}^{E_{\rm max}} d E_\tau \, \displaystyle\frac{d \Gamma}{d E_\tau}}
\end{equation}
and the central moments 
\begin{equation}
\label{eq:lept-energy-centr-mom-def}
\overline M_\tau^n \equiv 
\langle (E_\tau - \langle E_\tau \rangle)^n \rangle_{E_\tau > E_{\rm cut}},
\end{equation}
where $E_{\rm cut}$ denotes the energy cut of $\tau$-lepton and $E_{\rm max}$ is its maximal value.

\subsection{Numerical analysis and results}
\begin{table}[ht]
\begin{center}
\begin{tabular}{|c|c|l|c|}
\hline
Parameter & Value & Units  & Source \\
\hline
$m_b^{\rm kin}$ 			& $4.561 \pm 0.020$ 		& GeV 		&  \\
$m_c^{\rm kin}$ 			& $1.092 \pm 0.020$			& GeV 		&  \\
$\mu_\pi^2$ 				& $0.464 \pm 0.067$			& GeV$^2$ 	&  \\
$\mu_G^2$ 					& $0.333 \pm 0.061$			& GeV$^2$	&  \\
$\rho_D^3$ 					& $0.175 \pm 0.040$			& GeV$^3$	&  \cite{Alberti:2015} \\
$\rho_{LS}^3$				& $ -0.146 \pm 0.096 $		& GeV$^3$	&  \\
$V_{cb}\times 10^{-3}$ 		& $ 42.04 \pm 0.67 $		&  			&  \\
$\alpha_s$ 					& $0.218 \pm 0.018$			& 			&  \\
\hline
$G_F$ 						& $1.16637 \times 10^{-5}$ & GeV$^{-2}$& 	\\
$m_\tau$					& $1.777$				    & GeV		& \cite{PDG} 		\\
$\tau_{B^+}$				& $1.638$					& ps		& 	\\
$\tau_{B^0}$				& $1.520$					& ps		& 	\\
\hline
\end{tabular}
\caption{The values of the parameters involved in the decay width given in kinetic scheme.
The corresponding matrix of the correlations between the parameters  
can be found in  \cite{Alberti:2015}}
\label{tab:num-input}
\end{center}
\end{table}

We evaluate the rate and the moments in the kinetic scheme. To this end, we re-write the 
pole mass in (\ref{eq:decay-width-res}) in terms of the kinetic mass, using the one loop relation 
from \cite{Benson:2003kp}
\begin{equation}
m_{Q}^{\rm pole} = m_{Q}^{\rm kin} (\mu) 
\left( 1 + r_Q (\mu) \frac{\alpha_s}{\pi}  \right) 
\label{eq:mQ-pole-to-kin}
\end{equation}
with $Q = b$ or $c$ and auxiliary coefficient $r_Q$:
\begin{equation}
r_Q (\mu) = \frac{4}{3} C_F \frac{\mu}{m_{Q}^{\rm kin} (\mu)}
\left(1 + \frac{3}{8} \frac{\mu}{m_{Q}^{\rm kin} (\mu)} \right).
\label{eq:rQ}
\end{equation}
Inserting (\ref{eq:mQ-pole-to-kin}) into 
(\ref{eq:decay-width-res}) allows us to absorb parts of the one-loop 
QCD corrections into the mass definition.  In a similar way as for the light leptons, the 
remaining corrections are small and thus allow us a precise prediction. 

\begin{figure}[t]
\begin{center}
\includegraphics[scale=0.55]{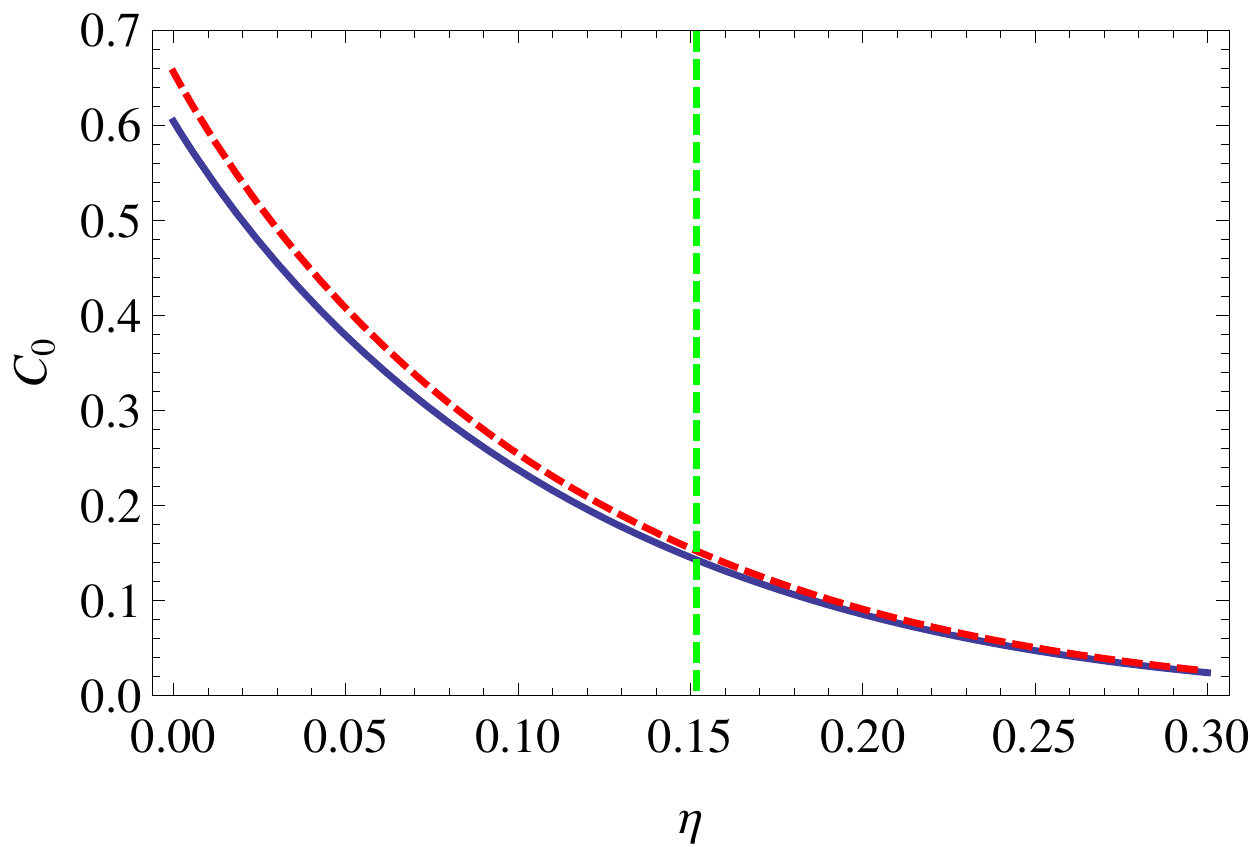}
\includegraphics[scale=0.55]{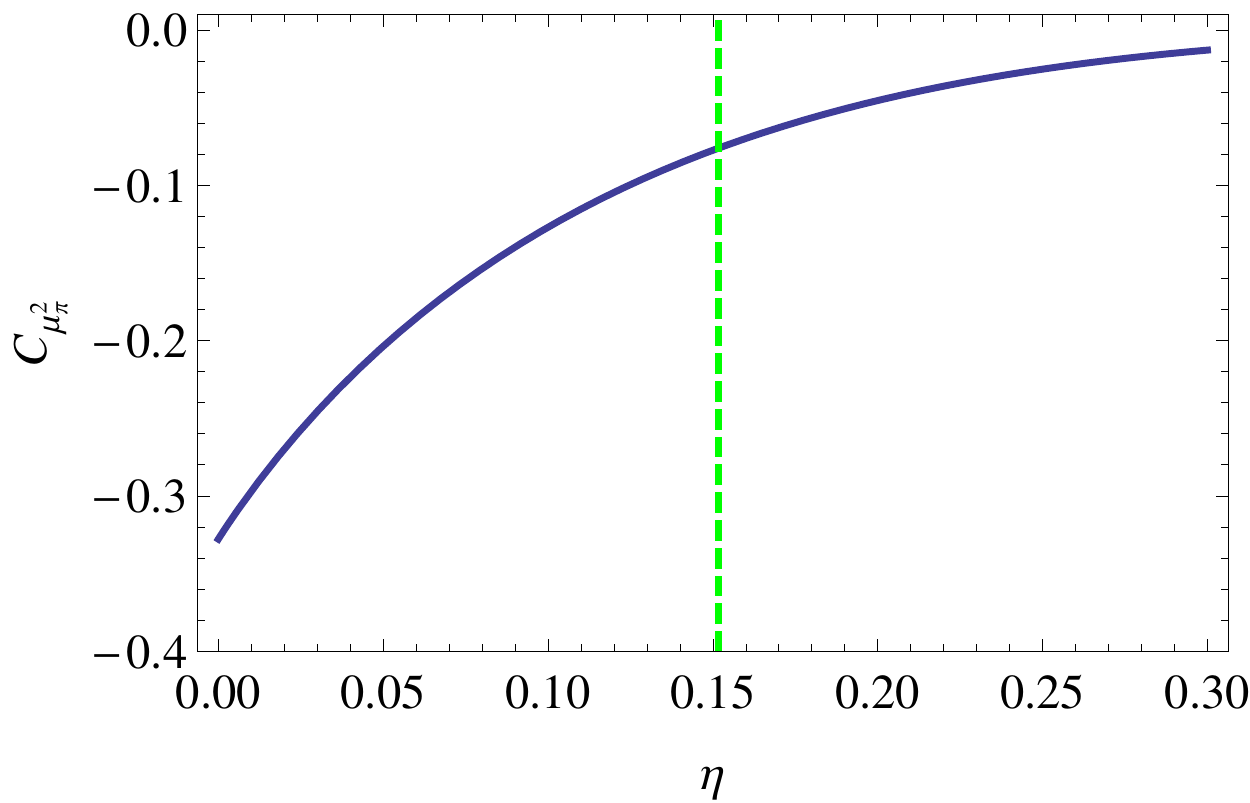}
\vspace*{5mm}

\includegraphics[scale=0.55]{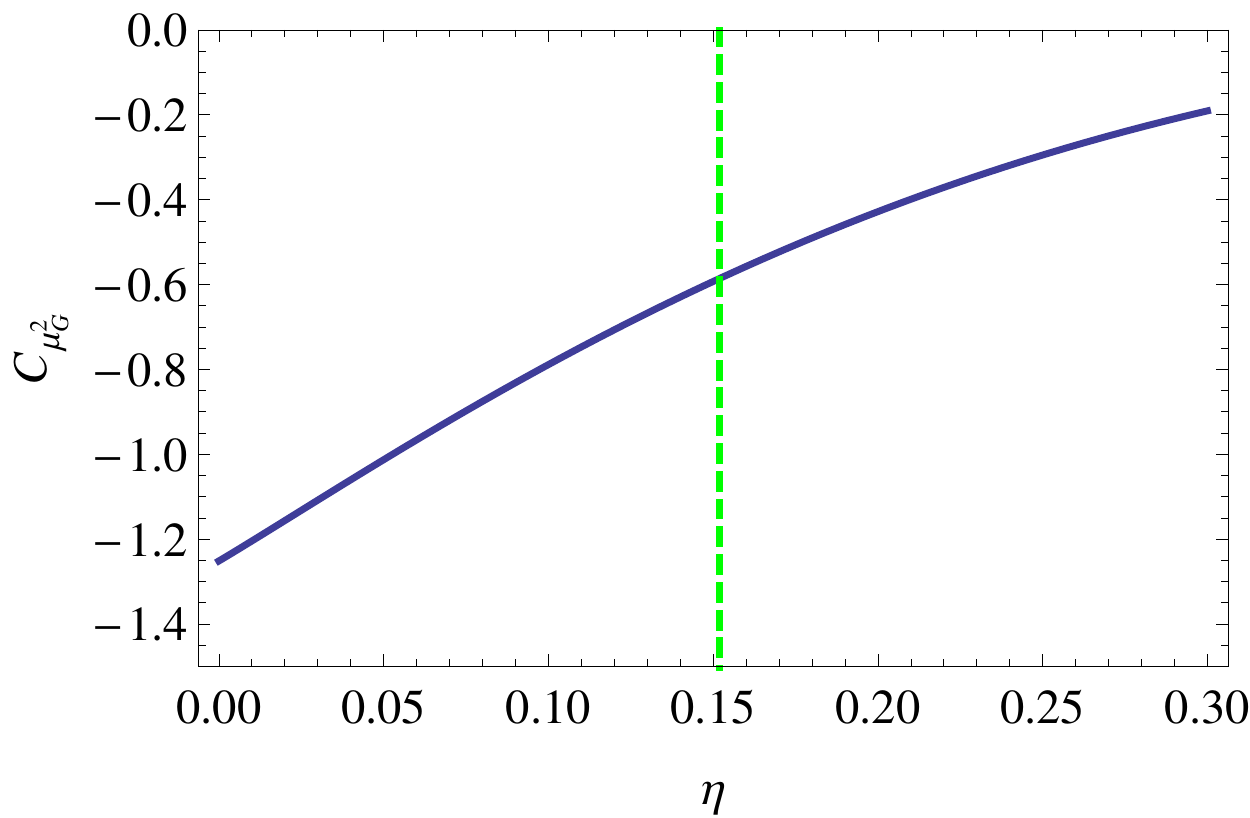}
\includegraphics[scale=0.55]{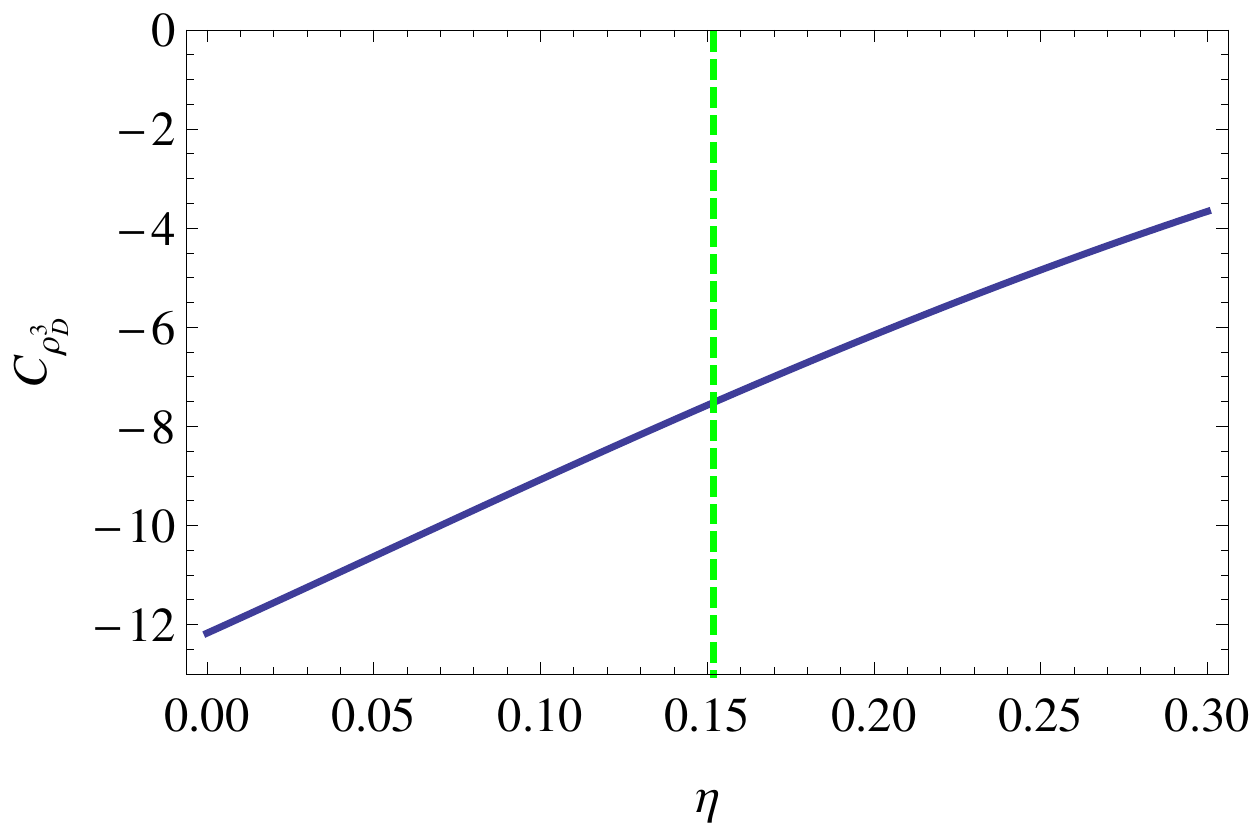}
\caption{Dependence of  $C_0, C_{\mu_\pi^2}, C_{\mu_G^2}, C_{\rho_D^3}$
on the parameter $\eta$, with $C_0 = C_0^{(0)} + \alpha_s / \pi \, C_0^{(1)}$.
In the left top plot the red dashed curve corresponds to $C_0^{(0)}$
and the blue solid one represents $C_0$ including the radiative correction.
The vertical dashed lines correspond to the central value of the parameter 
$\eta$ used in our analysis. All curves are plotted for central values of the input parameters.}
\label{fig:Coef-eta}
\end{center}
\end{figure}

The numerical values of the parameters used in our analysis are given in Tab.~\ref{tab:num-input}.  
For a simple comparison to the massless case we show the dependence of the 
coefficients $C_0, C_{\mu_\pi^2}, C_{\mu_G^2}, C_{\rho_D^3}$ 
in the kinetic scheme on the mass of the $\tau$ lepton in Fig.~\ref{fig:Coef-eta}.

\begin{table}[ht] 
\begin{center}
\begin{tabular}{|l|c|}
\hline
Accuracy & ${\rm Br} (B^+ \to X_c \tau^+ \nu_\tau) [\%]$ \\
\hline
LO 							  & $3.06 \, \pm  0.12$ \\
LO $ + \, 1/m_b^2$		 	  & $2.84 \, \pm  0.11$ \\
LO $ + \, 1/m_b^2 + 1/m_b^3$ & $2.56 \, \pm  0.09$ \\
NLO							  & $2.87 \, \pm  0.12$ \\
NLO $ + \, 1/m_b^2$ 		  & $2.65 \, \pm  0.10$ \\
\hline
NLO $ + \, 1/m_b^2 + 1/m_b^3$ & $2.37 \, \pm 0.08$ \\
\hline
\end{tabular}
\caption{
Values of the branching fraction of the inclusive $B^+ \to X_c^0 \tau^+ \nu_\tau$ decay 
depending on the different kinds of perturbative and power corrections included there.
The last row represents our final prediction for this process. Here the following value of 
charged $B$-meson life time $\tau_B^+ = 1.638$ ps is used \cite{PDG}.
In order to get results for neutral mode $B^0 \to X_c^- \tau^+ \nu_\tau$
it is sufficient to multiply values given in table by factor $\tau_B^0/ \tau_B^+ \approx 0.928$
\cite{PDG}.
}
\label{tab:Br-num}
\end{center}
\end{table}

Tab.~\ref{tab:Br-num} shows a breakdown of the various contributions for the total 
branching fraction, where we use the PDG values for the lifetimes. We can now
compute the total rate, and with the input of the measured lifetime we get 
for the branching fraction
of the inclusive $B^+ \to X_c \tau^+ \nu_\tau$ decay 
\begin{equation}
{\rm Br} (B^+ \to X_c \tau^+ \nu_\tau) =   (2.37 \, \pm 0.08) \%,
\label{eq:Br-incl-num-val}
\end{equation}
where the uncertainty appears due to a variation of the input parameters within
their intervals including the correlations between them.
The corresponding matrix of correlations between parameters shown in Tab.~\ref{tab:num-input} 
is not presented here and can be found in \cite{Alberti:2015}. 
The uncertainty in (\ref{eq:Br-incl-num-val}) includes also an estimate 
of the higher power contribution of order ${\cal O}(\Lambda_{\rm QCD}^4/m_b^4)$ 
where the relevant coefficient is conservatively assumed to be of order one. 
Moreover, we include the estimate of the contributions of the higher order
radiative corrections. We note that the corrections of order  
${\cal O} (\alpha_s ^2)$ have been computed in the on-shell scheme in \cite{Biswas:2009rb} 
and were found to be small. 
Thus we assume that the impact of the ${\cal O} (\alpha_s ^2)$ corrections 
in the kinetic scheme is within the quoted in (\ref{eq:Br-incl-num-val}) uncertainties.

Alternatively, we can compute the ratio 
$R (X_c) = {\rm Br} (B^+ \to X_c \tau^+ \nu_\tau) / {\rm Br} (B^+ \to X_c \ell^+ \nu_\ell)$ 
for which we obtain
\begin{equation}
R (X_c) = 0.212 \pm 0.003.
\end{equation} 
Combining this with the recent world average,
${\rm Br} (B \to X_c \ell  \nu_\ell) = (10.65 \pm 0.16) \% $, quoted by HFAG \cite{hfag}, 
we can avoid the uncertainty in $V_{cb}$, and we thus find an even more precise prediction 
\begin{equation}
{\rm Br} (B^+ \to X_c \tau^+ \nu_\tau) = (2.26 \pm 0.05) \% \, , 
\label{eq:Br-incl-SM-2way}
\end{equation}
with a slightly smaller central value compared to (\ref{eq:Br-incl-num-val}), 
which is, however, within the $1\sigma$ range.
We note that the uncertainty of our result (\ref{eq:Br-incl-SM-2way}) 
is comparable with one in \cite{LigetiN2}. 
However, our analysis shows that the coefficient in front of $\rho_D^3$ 
is of the order of ten, similar to what is observed for the case of a massless lepton. The result of including the $1/m_b^3$ corrections is thus a significant shift of 
the central value compared to the analysis up to $1/m_b^2$ as the one presented 
in  \cite{LigetiN2}, see also Tab.~\ref{tab:Br-num}.

Our predictions % (\ref{eq:Br-incl-num-val}) 
are also consistent with the measurement 
of the inclusive branching fraction of the LEP admixture of bottom baryons \cite{PDG}
\begin{equation}
{\rm Br} (b\mbox{-admix}  \to X \tau^\pm \nu) = (2.41 \, \pm 0.23) \% \, .
\label{eq:LEP-data}
\end{equation}

\begin{figure}[t] 
\begin{center}
\includegraphics[scale=0.9]{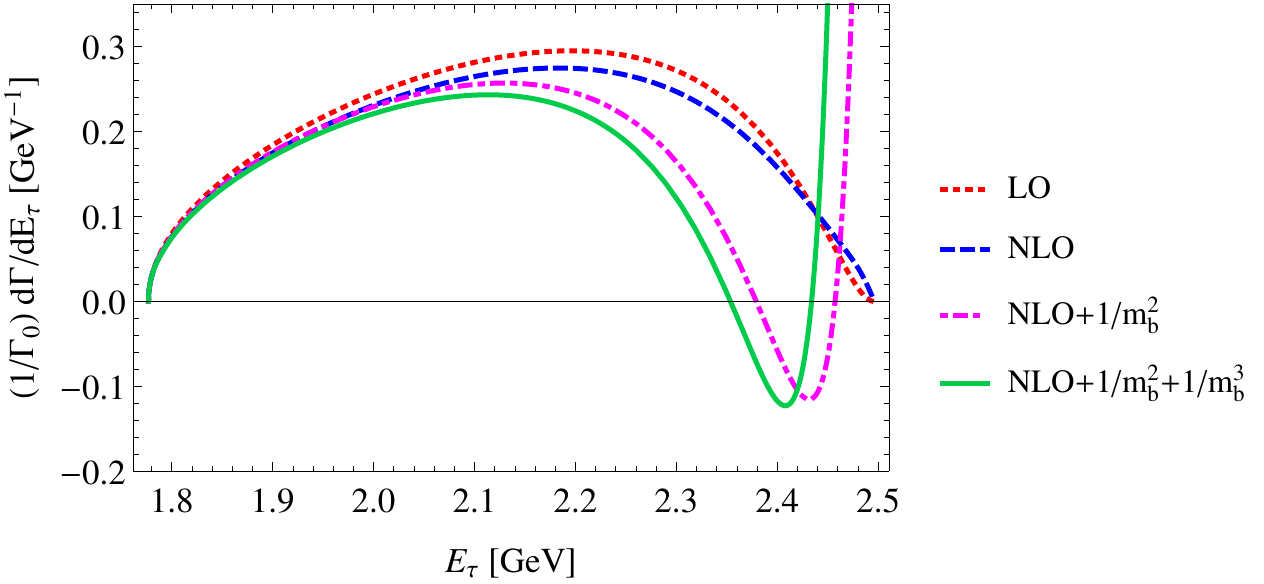}
\caption{$\tau$-lepton energy spectrum of the inclusive $\bar B \to X_c \tau \bar \nu_\tau$ decay.}
\label{fig:dGammadE}
\end{center}
\end{figure}

Moreover, we also show the resulting $\tau$-lepton energy distribution 
in Fig.~\ref{fig:dGammadE}. However, 
these curves cannot be interpreted on a point-by-point basis, since the OPE breaks down 
in the endpoint region. Note that this region is in fact larger than in the case 
of massless leptons due to the sizeable mass of the $\tau$ lepton.  
However, moments of these spectra can be interpreted in the $1/m_b$ expansion.
 
In \cite{LigetiN1} the authors also derived standard model predictions for the
$\tau$ energy distribution as well as dilepton invariant mass spectrum in the 
inclusive $B \to X_c \tau \bar \nu$ decay including $\Lambda_{QCD}^2/m_b^2$
and $\alpha_s$ corrections in the 1S mass scheme. In additions, they 
estimated the effects from shape functions in the endpoint region.
In our paper we focus on the $\tau$ energy distribution, including 
the $\Lambda_{QCD}^3/m_b^3$ corrections. 
In our analysis we use the kinetic scheme which explains some visible differences
between the shapes of the curves presented in Fig.~\ref{fig:dGammadE} of our paper 
and in Fig.~2 of~\cite{LigetiN1}. 

\begin{figure}[ht]
\begin{center}
\includegraphics[scale=0.6]{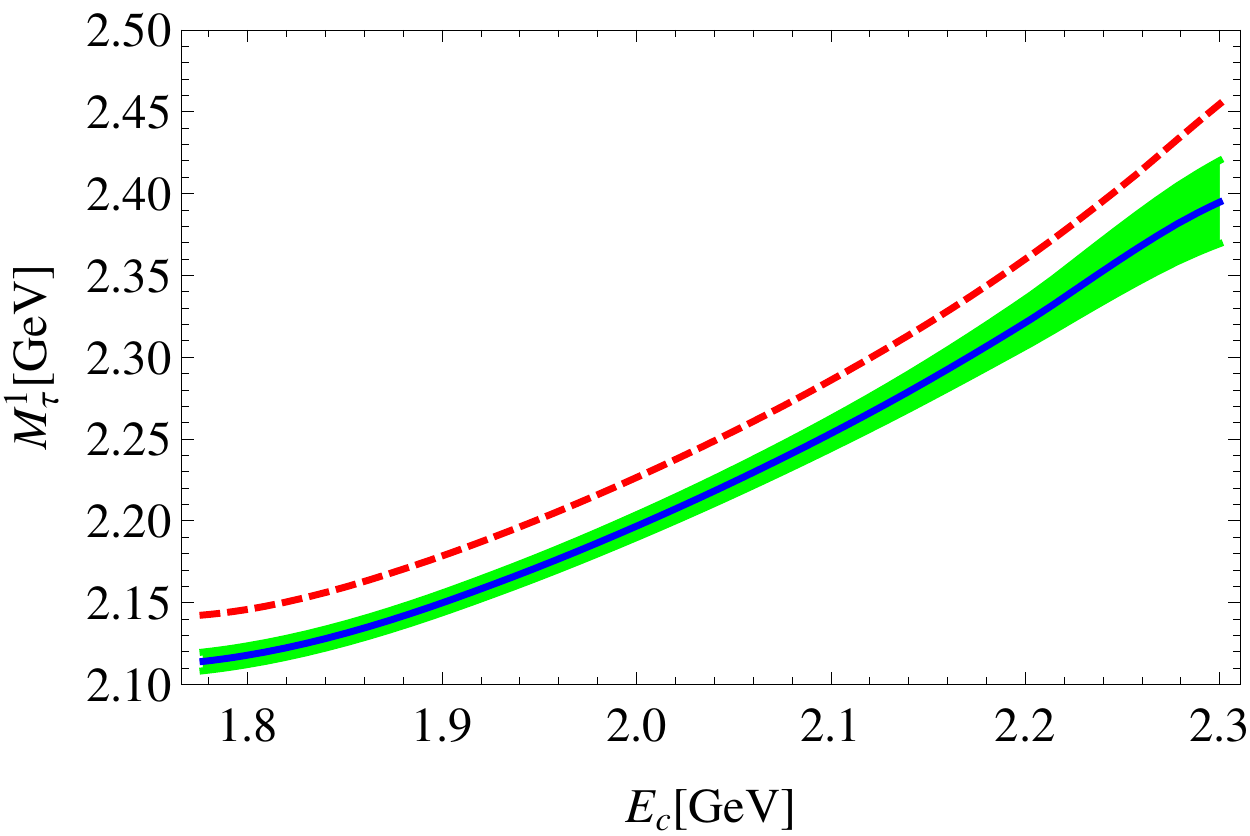}
\includegraphics[scale=0.6]{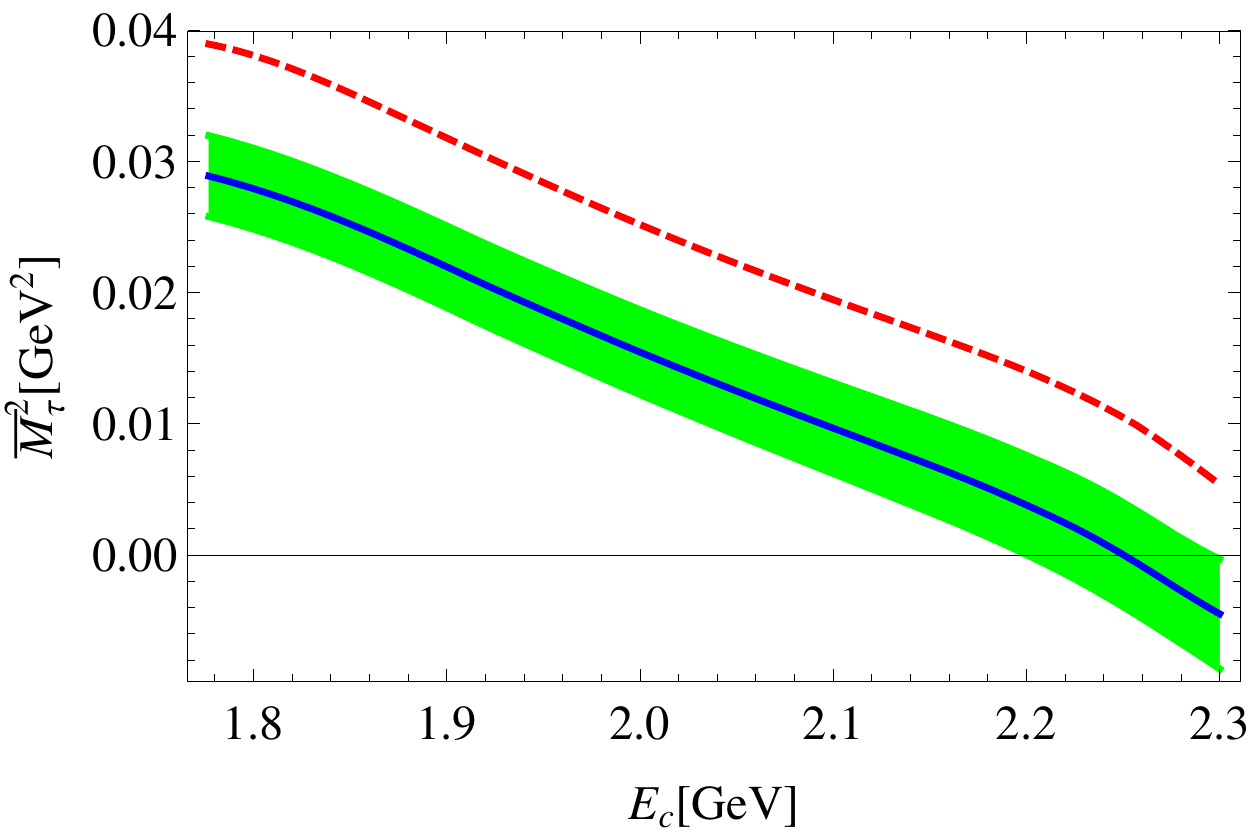}
\includegraphics[scale=0.6]{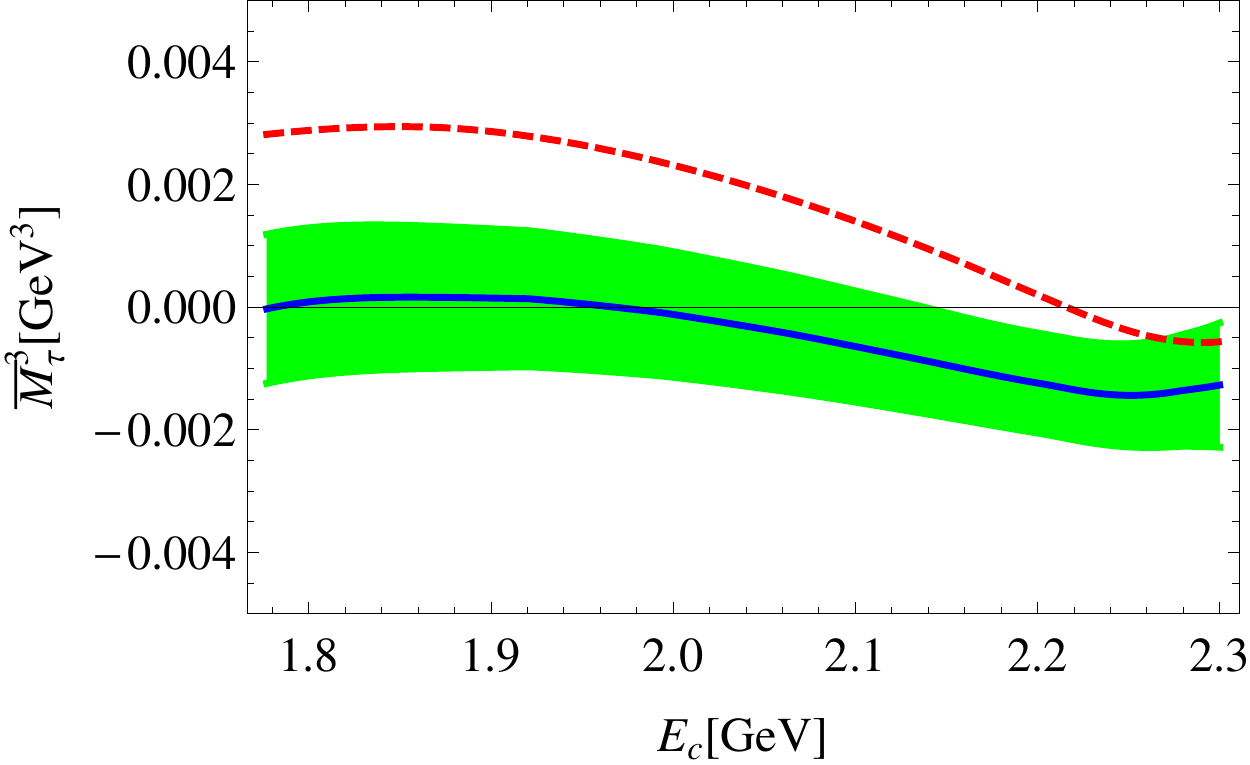}
\caption{Dependence of the moments of $\tau$-lepton energy spectrum on the value
of the cutoff energy $E_{\rm cut}$. The left plot corresponds to the first moment 
$M_\tau^1 (E_{\rm cut})$, the middle plot contains the second central moment 
$\overline M_\tau^2 (E_{\rm cut})$, the right plot contains the third central moment 
$\overline M_\tau^3 (E_{\rm cut})$. 
The solid curves with green shaded areas indicating uncertainties 
are the result of the full calculation, the dashed ones are without $1/m_b^3$ corrections.
}
\label{fig:moments}
\end{center}
\end{figure}

In the case of the light leptons the lepton energy distribution and
the relevant moments are the measurable observables.
However, for $\tau$ leptons, these observables will be more 
difficult to access, since the $\tau$ has to be reconstructed from its decay products.
Nevertheless, it is instructive to show the $\tau$-lepton energy moments defined 
by  eqs.~(\ref{eq:lept-energy-mom-def}) and (\ref{eq:lept-energy-centr-mom-def}) 
as a function of the cutoff energy $E_{\rm cut}$ 
for the sake of comparison with the light lepton case. 
We present our results in Fig.~\ref{fig:moments} and give the numerical results 
for several values of $E_{\rm cut}$ in Tab.~\ref{tab:moments}.
Once abundant data on this decay becomes available, appropriate inclusive observables 
have to be defined, which should take into account the decay of the $\tau$ lepton. 
The construction of such observables will be subject of future work.  

\begin{table}[t]\centering
\begin{tabular}{|l|c|c|c|}
\hline
Moment & $E_{\rm cut} = 1.8 \, {\rm GeV}$ & $E_{\rm cut} = 2.0 \, {\rm GeV}$ 
& $E_{\rm cut} = 2.2 \, {\rm GeV}$ \\
\hline
$M^1_\tau [{\rm GeV}] $ & 
$2.118 \pm 0.006$ & $2.197 \pm 0.007$ & $2.321 \pm 0.015$  \\
$\overline M^2_\tau [{\rm GeV}^2] $ & 
$0.028 \pm 0.003$ & $0.015 \pm 0.003$ & $0.004 \pm 0.004$  \\
$10^{3} \times \overline M^3_\tau [{\rm GeV}^3] $ & 
$0.08 \pm 1.21$ & $-0.12 \pm 1.02$ & $-1.24 \pm 0.82$  \\
\hline
\end{tabular}
\caption{The values of the moments of the $\tau$-lepton energy distribution
for three different values of the cutoff energy $E_{\rm cut}$.}
\label{tab:moments}
\end{table}
 
\section{The exclusive $\bar B \to D^{(*,**)} \tau \bar{\nu}$ decays}
Finally, we compare the inclusive result to the sum of identified exclusive states. 
The decays $\bar B \to D^{(*)} \tau \bar{\nu}$ into the two ground-state 
mesons $D$ and $D^*$ are described in terms of six form factors, some of which 
can be accessed in the corresponding decays into light leptons. However, due to the 
sizeable mass of the $\tau$ lepton there are two form factors which cannot be accessed 
from light-lepton data. For these one may make use of heavy quark symmetries to get 
at least an estimate. Still a quite precise prediction can be made due to the fact that 
the contribution of these form factors come with a suppression factor $m_\tau^2 / m_B^2$. 
Also the decays into the first orbitally excited mesons have been studied in the heavy mass 
limit. Using  QCD sum rules for the form factors appearing  in these processes one may get an 
estimate for these decays, which we shall generically denote as $\bar B \to D^{**} \tau \bar{\nu}$. 

In Tab.~{\ref{tab:excl-data}} we quote the recent SM predictions
for these processes, referring to \cite{Biancofiore:2013ki} for exclusive 
$\bar B \to D \tau \bar \nu$  and $\bar B \to D^* \tau \bar \nu$ decays and 
to \cite{Mannel:2015} and \cite{Bernlochner:2016bci} for $\bar B \to D^{(**)} \tau \bar \nu$. 
We note that the SM predictions for the exclusive channels 
$B^+ \to D^{(*,**)0} \tau^+ \nu$ imply
\begin{equation}
\label{eq:Br-excl-sum-numb}
{\rm Br} (B^+ \to D^0 \tau^+ \nu_\tau) +
{\rm Br} (B^+ \to D^{*0} \tau^+ \nu_\tau) +
\sum\limits_{D^{**}}{\rm Br} (B^+ \to D^{**0} \tau^+ \nu_\tau) = 
(2.14 \pm  0.16) \, \% \, .
\end{equation}
It is important to mention the recent paper \cite{Bigi:2016mdz} where 
the most precise prediction for $R(D) = 0.299 \pm 0.003$ was derived based on 
the combination of the experimental data and the result of the lattice calculation of 
the both $B \to D$ scalar and vector form factors \cite{Na:2015kha}.
%The ratios $R(D)$ and $R(D^*)$ are more precise determined both theoretically and
%experimentally. 
However, in our paper we focus on the calculation of the branching fraction 
of the inclusive decay and on the comparison with the corresponding branching fractions 
of the exclusive modes, and at this level the value given in eq.~(\ref{eq:Br-excl-sum-numb})
is sufficient for our purpose. 
From (\ref{eq:Br-excl-sum-numb}) one can see that 
the decays into the two ground state $D$ mesons already saturate the  
predicted inclusive rate to about 85\%, the lowest orbitally excited states add another 6\%,
leading to a saturation of the predicted  inclusive rate at a level of 90\%. 
This is in agreement with the expectation from the decays into light leptons, 
where the measured decay rates to the two ground state $D$ mesons saturate 
the measured inclusive rate at a level of about 72\%. 
Note that due to the sizeable $\tau$ lepton mass we expect a lesser degree of saturation 
for the light leptons, so the overall picture is very consistent. 

In Tab.~{\ref{tab:excl-data}} we also show the recent experimental data on 
$\bar B \to D \tau \bar{\nu}$ and  $\bar B \to D^* \tau \bar{\nu}$. 
We use the HFAG values for $R{(D)}$ and $R(D^*)$ and combine them with the PDG values 
for the branching ratios with light leptons to get the branching ratios for the semitauonic decays.  
Summing the experimental values for branching ratios into the two ground state $D$ mesons, 
we find an indication that these two decays alone already over-saturate 
the predicted inclusive rate, however, only at a level of $2 \sigma$. 
We take this as an indication of an inconsistency, which needs to be clarified.

\begin{table}[ht]
\begin{center}
\begin{tabular}{|c||c|c|} 
\hline
Mode & Theory (SM)  &  Experiment (HFAG + PDG) \\
\hline
${\rm Br} (B^+ \to D^0 \tau^+ \nu_\tau)$ 
& $(0.75 \pm 0.13) \, \%$  & $(0.91 \pm 0.11) \, \% $ \\ 
${\rm Br} (B^+ \to D^{*0} \tau^+ \nu_\tau)$ 
& $(1.25 \pm 0.09) \, \% $ & $(1.77 \pm 0.11) \, \% $ \\
\hline
${\rm Br} (B^+ \to (D^{0}+D^{*0}) \tau^+ \nu_\tau)$ & 
$(2.00 \pm 0.16) \, \% $ & 
$(2.68 \pm 0.16) \, \% $ \\
\hline
$\sum\limits_{D^{**}} {\rm Br} (B^+ \to D^{**0} \tau^+ \nu_\tau)$ & 
$(0.14 \pm 0.03) \, \% $ & --- \\ 
\hline
\end{tabular}
\caption{SM predictions and experimental results concerning the branching
fraction of $B^+ \rightarrow D^{(*,**)0} \tau^+ \nu_\tau$ decays.
In the second column the SM predictions of \cite{Biancofiore:2013ki}
and \cite{Bernlochner:2016bci} are presented and
in the third column the values extracted from combined data
provided by HFAG \cite{hfag} and PDG \cite{PDG} are given.} 
\label{tab:excl-data}
\end{center}
\end{table}

\section{Discussion and Conclusions} 

The tension of the recent data for $R(D)$ and $R(D^*)$ with the theoretical predictions for these 
exclusive channels has been intensively discussed recently, including a
possible explanation through effects from ``New Physics'' (NP).  
However, as it has been noticed before, there is also information on the inclusive rate,
experimental as well as theoretical. 

On the theoretical side, the heavy quark expansion allows us to perform a precise calculation of the 
inclusive semitauonic decay rates as well as of spectral moments. In the present paper we performed 
this calculation up to and including term at order $1/m_b^3$, thereby improving the existing calculations by one order in  the $1/m_b$ expansion. 
On the experimental side we have a measurement of the inclusive rate 
from LEP  which, however, is not precise enough to allow for a stringent test.

There have been various attempts to explain the tension in $R(D)$ and $R(D^*)$ 
in terms of different NP scenarios. We do not go into a detailed discussion of 
all possible scenarios, we rather parametrize the effects of NP by a simple extension 
of the effective Hamiltonian 
\begin{equation}
{\cal H}_{\rm NP}  = \frac{G_F V_{cb}}{\sqrt 2} 
\left(\alpha \, O_{V+A} +\beta \, O_{S-P} \right)
\label{eq:Hw-NP}
\end{equation}
with  the new operators  
\begin{eqnarray}
O_{V+A}  & = &  \left( \bar c \gamma_\mu (1 + \gamma_5) b \right) 
\left(\bar \tau \gamma^\mu (1 - \gamma_5) \nu \right), 
\label{eq:operators} \\
O_{S-P} & = &  \left( \bar c (1 - \gamma_5) b \right) 
\left(\bar \tau (1 - \gamma_5) \nu \right) 
\nonumber
\end{eqnarray}
and the dimensionless couplings $\alpha$ and $\beta$.  Our main motivation is to study the effect of (\ref{eq:Hw-NP}) on the 
inclusive rate on the basis of this example. 

We may discuss this effective low-energy interaction in the context of a standard-model effective theory (SMEFT) with linear realization of the Higgs field.  
It is interesting to note that the above operator structures  cannot be obtained at the leading order of the SMEFT expansion, since we insist on having lepton-universality violation. 
At dimension 6 we can write
\begin{equation}
P_{V+A}    =   \left( \bar c_R \gamma_\mu  b_R \right) (\phi^\dagger (i D^\mu) \phi)
\end{equation}
where $\phi$ is the SM Higgs doublet field. After spontaneous symmetry breaking, 
this operator generates an anomalous coupling of the $W$ to the right handed $b \to c$ current. 
Since the SM coupling to the $W$ is lepton universal, the insertion of this operator 
into the SM Lagrangian would lead to a lepton-universal effect. 
Thus we would need to combine this with another new-physics operator 
which will have dimension six and which generates a lepton-universality violating coupling of 
the $W$ to the left handed $\tau \to \nu$ current. 
Upon integrating out the $W$ boson, the combination of the two dimension-six operators
generates the same effects as the dimension eight operators we shall discuss now.   
In fact, writing the left-handed $SU(2)_L$ doublets as $L$ for the leptons and  $Q$ for the quarks,
we can construct the relevant $SU(2)_L \times U(1)_Y$ invariant operators of dimension eight
\begin{eqnarray}
O_{V+A}^\prime  & = &  \left( \bar c_R \gamma_\mu  b_R \right)
\left( (\bar L \cdot \phi^\dagger)
\gamma^\mu (\tilde{\phi} \cdot L ) \right),  \\
O_{S-P}^\prime  & = &  \left( \bar c_R (\phi^\dagger \cdot Q)  \vphantom{\tilde\phi}  \right) \left(\bar \tau_R (\tilde{\phi}^\dagger \cdot L)   \right),
\end{eqnarray}
where $\tilde\phi$ is the charge conjugate Higgs field. 
Once the Higgs field acquires its VEV
$$
\langle \phi \rangle = \frac{1}{\sqrt 2} \left(\begin{array}{c} 0 \\ v \end{array} \right), 
\quad 
\langle \tilde\phi \rangle = \frac{1}{\sqrt 2} \left(\begin{array}{c} v \\ 0 \end{array} \right),
$$
we obtain $8 \, O_{V+A}^\prime  = v^2 O_{V+A}$  and $8 \, O_{S-P}^\prime  = v^2 O_{S-P}$.
To this end, we infer that within the SMEFT power counting we have 
\begin{equation}
\alpha, \beta = {\cal O} \left( \frac{v^4}{\Lambda_{NP}^4} \frac{1}{V_{cb}} \right).
\label{eq:SMEFT-count}
\end{equation}   

However, the purpose of our simple ansatz (\ref{eq:Hw-NP}) is not a sophisticated analysis of NP effects, rather we 
want to study the effect of NP on the inclusive rate. 
It is a straightforward exercise to add (\ref{eq:Hw-NP}) to the effective Hamiltonian of the SM and to re-compute the 
exclusive decay rates for  ${\rm Br}_{\rm NP}  (B \to D^{(*)} \ell \nu_\ell)$ 
including the NP effects, using the 
heavy-quark limit for the form factors (see eg. \cite{Biancofiore:2013ki}).  
Assuming that there is no effect in the decays into light leptons, 
the resulting expressions for $R(D)$ and $R(D^*)$ are quadratic forms 
in the parameters $\alpha$ and $\beta$. 
We define the corresponding NP ratios $R_{\rm NP} (D^{(*)})$ by 
\begin{equation}
R_{\rm NP} (D^{(*)}) = 
\frac{{\rm Br}_{\rm NP} (\bar B \to D^{(*)} \tau \nu_\tau)}
{{\rm Br}  (\bar B \to D^{(*)} \ell \nu_\ell)}.
\label{eq:RD-NP-def}
\end{equation} 

The values of parameters $\alpha$ and $\beta$ are extracted by  requiring    
consistency with the corresponding experimental data on $R (D)$ and $R (D^*)$. 
Our ansatz is designed to describe both $R (D)$ and $R (D^*)$ simultaneously,
and a fit yields  
\begin{equation}
\alpha = -0.15 \pm 0.04, \quad \beta = 0.35 \pm 0.08  
\label{eq:alpha-beta-sol}
\end{equation}
for the parameters $\alpha$ and $\beta$. 
Note that there is a second solution, 
which exhibits destructive interference with the SM contribution. 
This solution yields a smaller (in comparison with the first scenario) value for 
the inclusive rate, which is in tension with the measurement of the 
sum of the branching fractions of the exclusive 
$B^+ \to D^0 \tau^+ \nu_\tau$ and $B^+ \to D^{0 *} \tau^+ \nu_\tau$ decays.  
It is interesting to note that the values (\ref{eq:alpha-beta-sol})
obtained in our fit are not in conflict with the above SMEFT discussion since
putting $v = $ 250 GeV, $\Lambda_{\rm NP} \sim 1$ TeV, $V_{cb} \simeq 0.041$  
in (\ref{eq:SMEFT-count}) yields $\alpha, \beta \sim 0.1$. 

It is worthwhile to point out a subtlety in the extraction of 
the parameters $\alpha$ and $\beta$ from the exclusive decays. 
The experimental analysis of $R(D)$ and $R(D^*)$ assumes the SM shapes for the 
kinematic distributions, which are used to extract e.g. efficiencies. 
However, including the NP operators  (\ref{eq:operators}) will change the shapes 
of the spectra, and hence the extracted values could shift. 
As in most other NP analyses we assume that this is only a small effect; 
a full analysis of this is clearly beyond the scope of this paper.  

We are now ready to study the impact of this NP model on the inclusive rate 
by including the NP operators (\ref{eq:operators}) into the calculation. 
Inclusion of NP modifies the parametrization (\ref{eq:decay-width-res}), which becomes 
\begin{equation}
\Gamma_{\rm NP} = \Gamma_{\rm SM} + 
\Gamma_0 \left[ A_1 \, \alpha + A_2 \,\alpha^2 + C_{12} \, \alpha \beta  + B_1 \, \beta + B_2 \, \beta^2 \right],
\label{eq:Gamma-NP}
\end{equation}  
with coefficients $C_0, A_1, A_2, B_1, B_2, C_{12}$ depending on parameters 
$\rho = m_c^2/m_b^2$ and $\eta = m_\tau^2 / m_b^2$,  
and  $\Gamma_{\rm SM} $ is the expression given in (\ref{eq:decay-width-res}). 

\begin{figure}[t]\center
\includegraphics[scale=0.7]{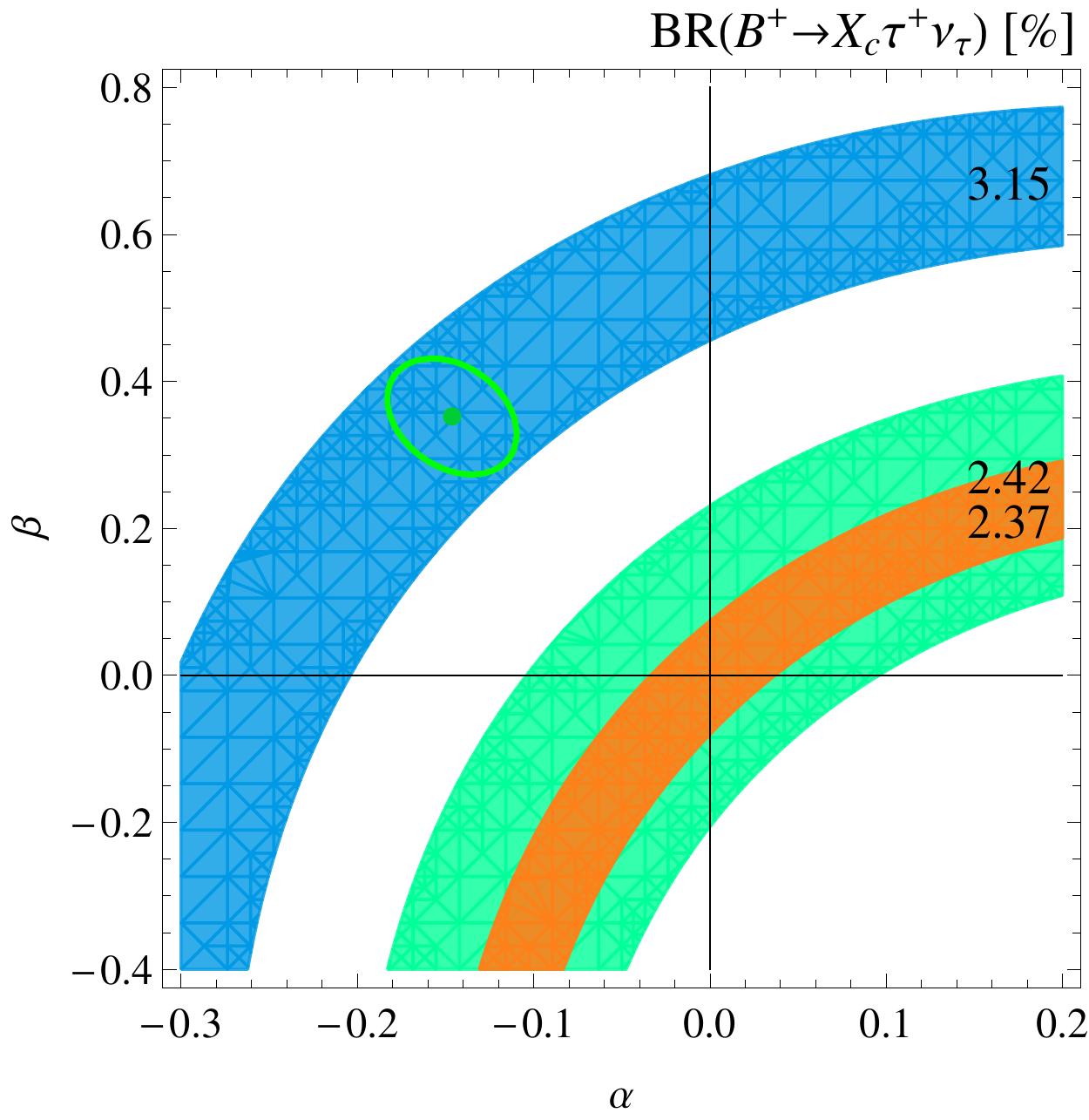}
\caption{Contour plot for the branching fraction of $B^+ \to X_c \tau^+ \nu_\tau$ 
(\ref{eq:Gamma-NP}) as function of $\alpha$ and $\beta$.
The green dot together with the ellipse indicate the best fit value and one-sigma range
of the parameters $\alpha$ and $\beta$
extracted from $R(D)$ and $R(D^*)$, see  (\ref{eq:alpha-beta-sol}).
The shaded bands indicate the one-sigma intervals of 
${\rm Br} (B^+ \to X_c \tau^+ \nu_\tau)$:
the red band is our SM prediction (\ref{eq:Br-incl-num-val}),
the green area represents the LEP measurement (\ref{eq:LEP-data}),
and the blue band is our prediction for inclusive $B^+ \to X_c \tau^+ \nu_\tau$
decay including contribution from NP (specified in Tab.~\ref{tab:summary}).}
\label{fig:Incl-Br-NP}
\end{figure}

In Fig.~{\ref{fig:Incl-Br-NP}} we show the dependence of the inclusive rate
on the parameters $\alpha$ and $\beta$.
The green dot together with the ellipse indicate the best fit value and one-sigma range,
respectively, of the parameters $\alpha$ and $\beta$ (\ref{eq:alpha-beta-sol})
extracted from $R(D)$ and $R(D^*)$.
The shaded bands indicate the one-sigma intervals for ${\rm Br} (B^+ \to X_c \tau^+ \nu_\tau)$:  
the red band is our SM prediction (\ref{eq:Br-incl-num-val}),
the green area represents the LEP measurement (\ref{eq:LEP-data}),
and the blue band is our prediction for inclusive $B^+ \to X_c \tau^+ \nu_\tau$
decay including the contribution (\ref{eq:Hw-NP}) from NP (specified in Tab.~\ref{tab:summary}).
The error estimate of the latter value contains also the uncertainties
from $\alpha$ and $\beta$, including the correlation between them. 
The NP prediction brings the inclusive rate  into agreement with the data on exclusive decays, 
but is now in visible tension with the LEP data.

Recently, the constraints on NP in the $b \to c \tau \bar{\nu}$ 
transition from the tauonic $B_c$ decay have been discussed, 
however, with a different ansatz for the NP operators 
\cite{Grinstein,Celis:2016azn,Li:2016vvp}.  
In fact, adding the new physics contribution (\ref{eq:Hw-NP}) 
yields a modification of the decay rate $B_c \to \tau \bar{\nu}$, which reads
\begin{equation}
\Gamma (B_c \to \tau \bar{\nu}_\tau) = \frac{M m_\tau^2 f_{B_c}^2 G_F^2 |V_{cb}|^2}{8 \pi}
 \left( 1- \frac{m_\tau^2}{M^2} \right)^{\! 2} 
\left| 1 - \alpha - \frac{M^2}{m_\tau (m_b + m_c)} \,  \beta \right|^2 , 
\label{eq:Bc-lept-decay}
\end{equation}  
where $M$ is the mass of the $B_c$ meson and $f_{B_c}$ is its decay constant, defined in the usual way. 
It has been pointed out in \cite{Grinstein} that  even relatively small 
values of $\beta$ may have a significant effect in the decay rate, 
since the pre-factor $ M^2 / (m_\tau (m_b + m_c)) \sim 4$ 
enhances the contribution of $O_{S-P} $. 

Using the parametrization (\ref{eq:Hw-NP}) together with our fit values  
implies a reduction of the tauonic branching fraction for the $B_c$ compared to the SM, 
since the extracted value of $\beta = 0.35$ is positive and yields in combination with 
the corresponding pre-factor the relative contribution of order $\sim 1$ 
but with a opposite sign compared to $(1 - \alpha)$ contribution, as one can see from 
(\ref{eq:Bc-lept-decay}). We conclude that  the width of leptonic 
$B_c \to \tau \bar \nu_\tau$ decay including our parametrization of NP 
is not in tension with the measured $B_c$ lifetime. 

Thus we arrive at a different conclusion compared to \cite{Grinstein}. 
However, the reason is that we dropped the assumption that only the leading order 
in the SMEFT expansion is taken into account.  Thus, attributing a possible  NP effect 
leading to the $R(D^{(*)})$ puzzle to dimension-eight operators can lift the constraint 
obtained in \cite{Grinstein}. We have pursued a different purpose with this simple model, 
but this observation might deserve a more detailed analysis.

\begin{table}[t]
\begin{center}
\begin{tabular}{|c|c|c|c|}
\hline
& SM & NP & Experiment \\
\hline
${\rm Br} (B^+ \to D^0 \tau^+ \nu_\tau)$ & 
$(0.75 \pm  0.13)$ \% &
$0.93  $ \% &
$(0.91 \pm 0.11)$ \% \\
\hline
${\rm Br} (B^+ \to D^{*0} \tau^+ \nu_\tau)$ & 
$(1.25 \pm 0.09) $ \% &
$1.65  $ \% &
$(1.77 \pm 0.11)$ \% \\
\hline
${\rm Br} (B^+ \to X_c \tau^+ \nu_\tau)$ & 
$(2.37 \pm 0.08)$ \% &
$(3.15 \pm 0.19)$ \% &
$(2.41 \pm  0.23)$ \% \\
\hline
\end{tabular}
\caption{Summary of predictions for different mode of semitauonic $B$-meson decays in 
the framework of SM and including NP effects in comparison with relevant experimental data. NP predictions presented here correspond to
the first scenario for parameters $\alpha, \beta$ (\ref{eq:alpha-beta-sol}). We do not quote an uncertainty for the exclusive NP calculations; for fixed 
$\{\alpha, \beta\}$ the uncertainties are of the same size as the SM ones. }
\label{tab:summary}
\end{center}
\end{table}

\section*{Acknowledgements}
TM thanks Martin Jung for useful discussions. 
This work was supported by the Research Unit FOR 1873, funded by the German research foundation DFG. 
AR and FS acknowledge the support by a Nikolai-Uraltsev Fellowship of Siegen University.

\section{Appendix}

Here the explicit  analytic expressions of the coefficients
introduced in the $B \to X_c \tau \nu_\tau$ decay width (\ref{eq:decay-width-res}) 
as functions of dimensionless variables
$\rho$ and $\eta$ are given:

\begin{eqnarray}
C_0 & = &
\sqrt{R} \left[1 - 7 \rho - 7 \rho^2 + \rho^3 - (7 - 12 \rho + 7 \rho^2) \eta  
- 7 (1 + \rho) \eta^2 + \eta^3 \right] \\
& - & 
12 \left[\rho^2\,
  \ln \frac{(1 + \rho - \eta - \sqrt{R})^2}{4\rho} 
- \eta^2\,\text{ln}\frac{(1+\eta-\rho + \sqrt{R})^2}{4\eta}
- \rho^2 \eta^2 \, \ln \frac{(1-\rho-\eta-\sqrt R)^2}{4 \rho \eta} \right] \!,
\nonumber
\label{eq:C0-AE}
\end{eqnarray}

\begin{eqnarray}
C_{\mu_\pi^2} & = &
- \frac{\sqrt{R}}{2} \left[1 - 7 \rho - 7 \rho^2 + \rho^3 - (7 - 12 \rho + 7 \rho^2) \eta  
- 7 (1 + \rho) \eta^2 + \eta^3 \right] \\
& + & 
6 \left[\rho^2\,
  \ln \frac{(1 + \rho - \eta - \sqrt{R})^2}{4\rho} 
- \eta^2\,\text{ln}\frac{(1+\eta-\rho + \sqrt{R})^2}{4\eta}
- \rho^2 \eta^2 \, \ln \frac{(1-\rho-\eta-\sqrt R)^2}{4 \rho \eta} \right] \!,
\nonumber
\label{eq:Cmupisq-AE}
\end{eqnarray}

\begin{eqnarray}
C_{\mu_G^2} & = &
\frac{\sqrt{R}}{2} \left[ - 3 + 5 \rho - 19 \rho^2 + 5 \rho^3 
+ (5 + 28 \rho - 35 \rho^2) \eta - (19 + 35 \rho) \eta^2 + 5 \eta^3 \right] \\
& - & 
6 \left[\rho^2\,
  \ln \frac{(1 + \rho - \eta - \sqrt{R})^2}{4\rho} 
- \eta^2\,\text{ln}\frac{(1+\eta-\rho + \sqrt{R})^2}{4\eta}
- 5 \rho^2 \eta^2 \, \ln \frac{(1-\rho-\eta-\sqrt R)^2}{4 \rho \eta} \right]\! ,
\nonumber
\label{eq:CmuGsq-AE}
\end{eqnarray}

\begin{eqnarray}
C_{\rho_D^3} & = & 
\frac{2}{3\rho^2\sqrt{R}}\Big\{
\eta ^7 - \eta ^6 (5 \rho + 7) + \eta ^5 \left(6 \rho^2 + 22 \rho + 21 \right)
+ \eta^4 \left(\rho ^3 - 9 \rho^2 - 35 \rho - 35 \right)  \\ 
& + & \eta ^3 (-5 \rho^4 - 2 \rho ^3 -8 \rho^2 + 20 \rho + 35) +  
\eta^2 \left(3 \rho ^5 - \rho^4 - 4 \rho^3 + 18 \rho^2 + 5 \rho - 21 \right) 
\nonumber \\
& + & \eta (1 - \rho)^3 \left(2 \rho^3 + 6 \rho^2 +11 \rho + 7 \right) 
+ (\rho-1)^5 (\rho + 1)^2
\nonumber \\
& - & 
R \Big[\eta^5 - \eta^4 (3 \rho + 5) +\eta^3 \left(- 3 \rho^2 + 8 \rho +10 \right)
+ \eta^2 \left(37 \rho^3 + 27 \rho^2 - 6 \rho - 10 \right)
\nonumber \\
& + & \eta \left(32 \rho^4 - 18 \rho^3 - 9 \rho^2 + 5 \right)
- 4 \rho^5 + 10 \rho^4 - 3 \rho^3 - 15 \rho^2 + \rho - 1 \Big] \Big\}
\nonumber \\
 & + & 
 8 \, \Bigg\{\eta^2 (5 \rho^2 + \eta - 1)
 \, \ln \! \left[ \cfrac{(1-\rho-\eta - \sqrt{R})^2}{4\eta\rho} \right]
 - (\eta - 1) \, \ln \! \left[ \cfrac{(1+\rho-\eta-\sqrt{R})^2}{4\rho} \right] \Bigg\},
 \nonumber
\end{eqnarray}
where $R = \eta^2 - 2 \, \eta \, (\rho + 1) + (\rho - 1)^2$.

\end{document}